\documentclass{article}

\usepackage{arxiv}

\usepackage[utf8]{inputenc} % allow utf-8 input
\usepackage[T1]{fontenc}    % use 8-bit T1 fonts
\usepackage{hyperref}       % hyperlinks

\usepackage{booktabs}       % professional-quality tables
\usepackage{amsfonts}       % blackboard math symbols
\usepackage{nicefrac}       % compact symbols for 1/2, etc.
\usepackage{microtype}      % microtypography
\usepackage{lipsum}
\usepackage{graphicx}
\usepackage{amsmath}
\usepackage{amsthm}
\usepackage{amssymb}

\usepackage[super,comma,sort&compress]{natbib}
\usepackage{abstract}

\usepackage{lipsum}

\newtheorem{definition}{Definition}

\def\BibTeX{{\rm B\kern-.05em{\sc i\kern-.025em b}\kern-.08emT\kern-.1667em\lower.7ex\hbox{E}\kern-.125emX}}
    
\usepackage{comment}
\usepackage{graphicx,subcaption}
\usepackage{enumitem}

\title{A pipeline for integrated theory and data-driven modeling of genomic and clinical data}

\author{
Vineet K. Raghu\\
  Department of Computer Science\\
  University of Pittsburgh\\
  \texttt{vineet@cs.pitt.edu} \\
   \And
Xiaoyu Ge\\
  Department of Computer Science\\
  University of Pittsburgh\\
  \texttt{xig34@pitt.edu} \\
   \And
Arun Balajee\\
  Department of Computer Science\\
  University of Pittsburgh\\
  \texttt{abl55@pitt.edu} \\
   \And
Daniel J Shirer\\
  Department of Computer Science\\
  University of Pittsburgh\\
  \texttt{djs134@pitt.edu} \\
   \And
Isha Das\\
  Department of Computational and Systems Biology\\
  University of Pittsburgh\\
   \And
Panayiotis V. Benos\\
  Department of Computational and Systems Biology\\
  University of Pittsburgh\\
  \texttt{benos@pitt.edu} \\
   \And
Panos K. Chrysanthis\\
  Department of Computer Science\\
  University of Pittsburgh\\
  \texttt{panos@cs.pitt.edu} \\
}

\begin{document}

\maketitle
%
% The abstract is a short summary of the work to be presented in the article.
\begin{abstract}
High throughput genome sequencing technologies such as RNA-Seq and Microarray have the potential to transform clinical decision making and biomedical research by enabling high-throughput measurements of the genome at a granular level. However, to truly understand causes of disease and the effects of medical interventions, this data must be integrated with phenotypic, environmental, and behavioral data from individuals. Further, effective knowledge discovery methods that can infer relationships between these data types are required. In this work, we propose a pipeline for knowledge discovery from integrated genomic and clinical data. The pipeline begins with a novel variable selection method, and uses a probabilistic graphical model to understand the relationships between features in the data. We demonstrate how this pipeline can improve breast cancer outcome prediction models, and can provide a biologically interpretable view of sequencing data.
\end{abstract}
%
% Keywords. The author(s) should pick words that accurately describe the work being
% presented. Separate the keywords with commas.

%%TODO Include Isha's results via a couple lines
\section{Introduction}
Since the advent of high-throughput sequencing assays, a number of modeling approaches have been developed to predict patient outcome from genomic data \cite{hira2015review,cun2012prognostic}. To understand the complex relationships between genomics and outcomes, the genomic data are often merged with clinical and demographic information in an "integrated" dataset. The reason standard machine learning models have been insufficient to model this data \cite{haury2011influence} are due to its unique complexities. Specifically, these data have highly correlated sets of variables (genes), and often consist of several orders of magnitude more variables than samples (high-dimensional). Lastly for biomedical research purposes, predictive power of models are important, but interpretability of models are equally crucial. Often, biomedical researchers aim to \emph{learn from their models} to identify promising new hypotheses or to prioritize future experiments, instead of solely aiming to predict outcomes accurately.

Probabilistic Graphical Models (PGM's) are an effective tool to build interpretable models \cite{koller2009probabilistic}. These models represent a dataset as a graph where nodes correspond to features and edges correspond to dependence relationships. Learning the structure of these models from data is a well-studied problem for continuous data and categorical data but not mixtures of both. Recently, several approaches have been proposed to model mixed data \cite{Sedgewick2016LearningSelection,raghu2018comparison,Lee2013LearningModels}. However, to utilize these approaches requires addressing some of the aforementioned difficulties inherent to the integrated datasets. Graphical model learning would be computationally intractable on genome scale data. Further, interpretation of a large graphical model is difficult unless single variables of interest are queried. Lastly, highly correlated data can result in the formation of disconnected cliques, impeding model accuracy \cite{lemeire2012conservative}. 

\begin{figure*}
\centering
  \includegraphics[width=\textwidth]{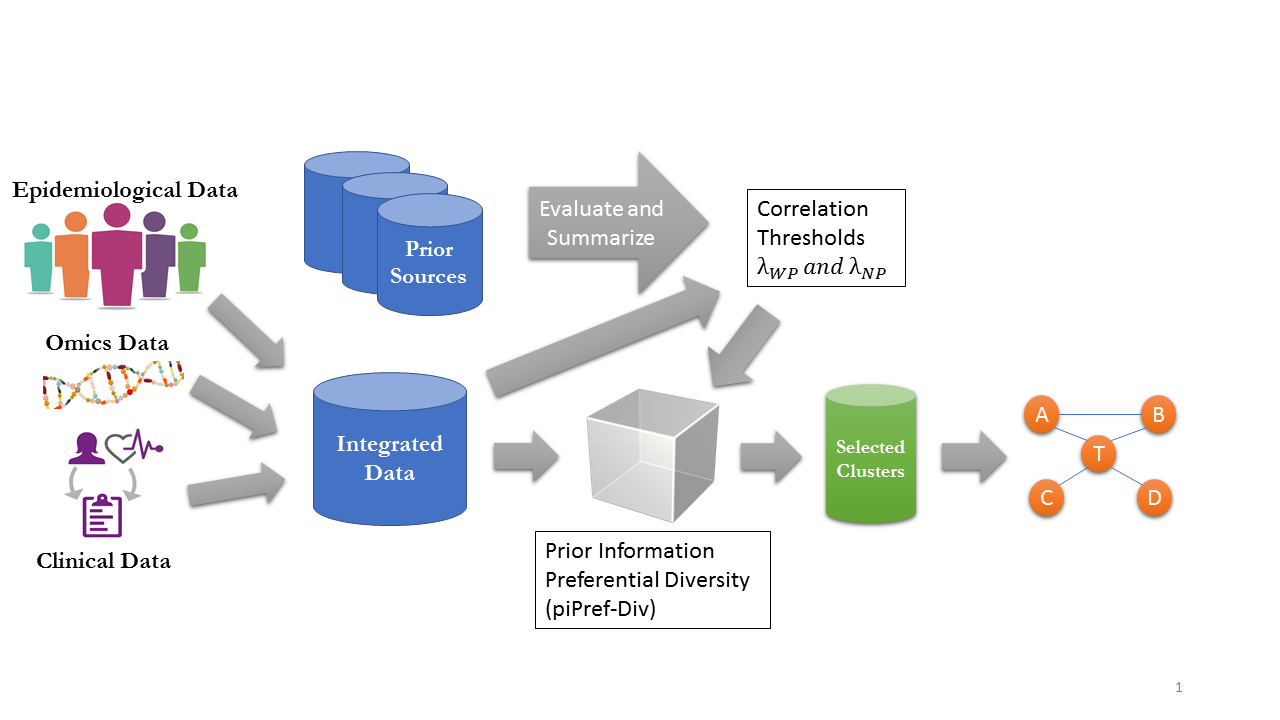}
  \caption{Pipeline proposed in this work to learn graphical model structure from mixed clinical and omics datasets.}
  \label{fig:teaser}
\end{figure*}

One way to address these issues is by selecting a subset of variables to model. The machine learning community refers to this problem as feature selection. There, the aim is to find the subset of features that best predicts a target variable of interest. Though some of these approaches are applied to integrated biomedical data, they still fail to address the aforementioned challenges. High correlation among features results in unstable prediction models, and harms interpretability of learned models \cite{Huang2015T-recs:Outcomes.}.

Using prior knowledge has been proposed as a way to address these difficulties \cite{cun2012prognostic,allahyar2015feral,taylor2009dynamic}. These sources allow a researcher to choose the most biologically plausible model among statistically equivalent modelsa \cite{Huang2015T-recs:Outcomes.,venet2011most}. However, many proposed methods have shown no significant benefit from using prior knowledge \cite{cun2012prognostic,staiger2013current}. Our hypothesis for this is twofold. First, external data sources need to be evaluated and weighted accordingly due to data provenance and context-specific information.  For example, a particular biological pathway may not be active in the context from which a genomic dataset was measured. Thus, this source of information should be downweighted in the final model. Second, multiple sources of prior information should be integrated to achieve consistent results.

This motivates our pipeline for modeling an integrated genomic and clinical dataset (Figure \ref{fig:teaser}). The first step is to measure the concordance between the data and each of the prior information sources, and weight the sources accordingly. This is based upon a prior knowledge evaluation method we recently developed for graphical structure learning \cite{manatakis2018pimgm}. Then, the information in the data and the prior knowledge are fused to select parameters for a feature selection method (Pref-Div). Finally, the clusters selected by Pref-Div are modeled using a graphical model structure learning algorithm to represent the dependencies between the clusters and outcome variables of interest.

Our specific contributions are as follows:
\begin{itemize}[noitemsep,topsep=0pt]
    \item A novel method for variable and cluster selection that combines a feature selection approach \cite{PrefDiv} with an approach to evaluate and integrate prior information \cite{manatakis2018pimgm}.
    \item An extensive evaluation of the variable selection approach on synthetic datasets
    \item A pipeline that combines variable selection with graphical modeling to represent inter-relationships between variables from mixed discrete and continuous datasets
    \item An evaluation of the pipeline against state of the art variable selection approaches in predictiing breast cancer outcome and subtyping from transcriptomic data.
\end{itemize}
%%Give a brief overview of the proposed pipeline and how we go from prior information sources to hyperparameter selection to features, to aggregates of features to graphical model 

%%Make a figure for this^^

\section{Related Work}

In this section, we give background information on feature selection methods for genomic data. Then, we discuss how prior knowledge has been incorporated. Finally, we discuss graphical model structure learning approaches for mixed datasets.

\subsection{Feature Selection in Genomics}
Feature selection aims to identify a subset of features in a dataset that together best predict a target variable \cite{guyon2003introduction}. The main purpose of feature selection is to improve model training efficiency and to prevent overfitting. In machine learning, feature selection approaches fall into three broad classes: filter methods, wrapper methods, and embedded methods \cite{haury2011influence}. Filter methods select features using univariate ranking scores such as a Wilcoxon test or a t-test between a covariate and a target variable. Wrapper methods use a predictive model like the Support Vector Machine to select a set of features that result in an accurate prediction model \cite{johannes2010integration}. Two popular wrapper methods are the recursive feature elimination and greedy forward search, which select the best feature to eliminate (or include, respectively) in a step-wise fashion. Finally, embedded methods are predictive models which select features automatically as part of the learning procedure. The most popular example of this is the LASSO regression method \cite{tibshirani1996regression}, which uses an $L_{1}$ norm penalty to shrink coefficients to zero in a linear regression. 

Recently, a study was performed investigating the performance of these techniques to predict breast cancer relapse from genomic data \cite{haury2011influence}. Overall, no method had significantly better accuracy than randomly choosing features and learning a classification model, and only filter based methods were more stable (insensitive to variations in the data). Since this was true even within a single dataset, it could only be attributed to the methods themselves. This suggests that tailored approaches are necessary to improve feature selection from omics data.

\subsection{Incorporating Prior Knowledge}
One way to improve these approaches is to use domain knowledge about the relationships between genes and between protein products. Three main sources of prior knowledge have been explored: gene ontology (GO) terms, protein-protein interaction networks (PPI's), and biological pathways \cite{hira2015review}. 

GO groups genes based on known biological functions (e.g. cell cycle or angiogenesis). Several approaches have leveraged GO terms as prior information to construct gene clusters \cite{pan2006incorporating,cheng2004knowledge,chen2009integrating}. The main drawback of these methods is the nature of GO terms. Not all genes belong to a functional group in the GO, and these methods chose to discard those genes. In addition, GO terms tend to define broad functional classes which are difficult to interpret.

PPI's are networks that encode protein interactions known to occur in normal cellular activity. Methods for gene selection have been built off of these networks, and they were reviewed and evaluated in \cite{cun2012prognostic}. Many of these approaches aim to either 1) group genes based on the edges in the network and penalize them together \cite{allahyar2015feral,wang2008hybrid,simon2013sparse,zhu2009network} or 2) use the network information to determine gene importance \cite{taylor2009dynamic,johannes2010integration}. Pathway based approaches are similar in principle, but use biological pathways which represent a module of the network that carries out a specific function. These are normally taken from a pathway database such as KEGG or I2D \cite{kanehisa2000kegg,brown2005online}. Biological pathway-based feature selection is a step-wise method that uses mutual information to the target variable as a scoring criterion, but does not choose multiple genes from the same area in the graph consisting of the union of all the pathways \cite{bandyopadhyay2010pathway}. In \cite{guo2005towards}, the authors attempt to construct a single feature for each pathway by aggregating information across multiple genes. A similar method is taken in \cite{alcaraz2017novo} except that the pathways are constructed using the data. Multiple studies have found no significant benefit in prediction accuracy over baseline methods; however, these methods do appear to give more biologically interpretable signatures \cite{cun2012prognostic,staiger2013current}.

%%%prior knowledge was thought to be a way to improve the performance, explain prior knowledge approaches (PPI, GO, etc.) -> no significant benefit in accuracy, but can improve stability and interpretability 

\subsection{Mixed Graphical Models}
For data exploration applications, graphical models enable a user to identify all direct associations for any variable of interest. Since genomic data is often integrated with clinical, demographic, and epidemiological data, in this work we focus upon approaches to learn undirected graphical models from mixed datasets: mixed graphical models (MGM). A MGMl parametrizes the joint distribution of a mixed dataset as a graph $G = \left(V,E\right)$, where $V$ is the set of variables and $E$ is the set of edges. In this type of model, an edge exists between two variables $X$ and $Y$ if $X$ and $Y$ are conditionally dependent given the rest of the variables in the data. 

Recently several methods have been proposed to learn MGM's. In \cite{tur2011learning}, the authors propose qp-graphs which can be estimated from high dimensional data.% The idea is to use a linear measure of association called limited-order correlations to remove edges from the model. These limited-order correlations can be estimated from small sample size data. 
However, this type of model assumes that there are no edges between categorical variables which is a limiting assumption for clinical data. Several authors have proposed using regression based methods to estimate the conditional dependencies among pairs of variables to infer the edges in the graph. In these approaches, each variable in turn is considered as the target variable, a regression is performed using all other variables as predictors, and edges are added to the model for all significant regressors. In, \cite{fellinghauer2013stable}, the authors use a random forest regression approach to rank edges for inclusion into a graphical model among mixed variables. In, \cite{yang2014mixed}, they assume that the conditional distributions of each type of variable come from the exponential family and use node-wise regression approaches to estimate the parameters of the model. Other authors have proposed similar techniques to the aforementioned methods \cite{cheng2016high,chen2014selection,friedman2008sparse}. 

Another way to estimate a MGM is the pseudolikelihood approach \cite{besag1975statistical}. This approach uses the product of conditional distributions of the variables as a consistent estimator of the true likelihood without computing the partition function. Then a gradient based optimization method can be used to find maximum pseudolikelihood estimates of the parameters. Lee and Hastie propose a MGMl that generalizes a popular continuous graphical model (Gaussian Graphical Model) and a popular discrete model (Markov Random Field) \cite{Lee2013LearningModels}. They demonstrate that using the pseudolikelihood approach shows better empirical performance than using separate regressions, and so we focus on this approach. 

The authors parameterize the joint distribution of the variables according to Equation \ref{EQ:Joint}.

%%%TODO Deal with duplicate numbering
\begin{align}
\begin{split}
\label{EQ:Joint}
p(x,y;\theta)  \propto  exp\bigg(\sum\limits_{s=1}^{p}\sum\limits_{t=1}^{p} -\frac{1}{2}\beta_{st}x_{s}x_{t} + \sum\limits_{s=1}^{p}\alpha_{s}x_{s} + \\ \sum\limits_{s=1}^{p}\sum\limits_{j=1}^{q}\rho_{sj}(y_{j})x_{s} +  \sum\limits_{j=1}^{q}\sum\limits_{r=1}^{q}\phi_{rj}(y_{r},y_{j})\bigg)
\end{split}
\end{align}

%
%\begin{equation}
%\label{EQ:Joint}
%p(x,y;\theta)  \propto  exp\bigg(\sum\limits_{s=1}^{p}\sum\limits_{t=1}^{p} -\frac{1}{2}\beta_{st}x_{s}x_{t} + \sum\limits_{s=1}^{p}\alpha_{s}x_{s} + \sum\limits_{s=1}^{p}\sum\limits_{j=1}^{q}\rho_{sj}(y_{j})x_{s} +  \sum\limits_{j=1}^{q}\sum\limits_{r=1}^{q}\phi_{rj}(y_{r},y_{j})\bigg)
%\end{equation}

where $\theta$ represents the full set of parameters,  $x_{s}$ represents the $s^{th}$ of $p$ continuous variables and $y_{j}$ represents the $j^{th}$ of $q$ discrete variables.  $\beta_{st}$ represents the edge potential between continuous variables $s$ and $t$, $\alpha_{s}$ represents the continuous node potential for $s$, $\rho_{sj}$ represents the edge potential between continuous variable $s$ and discrete variable $j$, and finally $\phi_{rj}$ represents the edge potential between discrete variables $r$ and $j$. This model has the favorable property that its conditional distributions are given by Gaussian linear regression and Multiclass Logistic Regression for continuous and discrete variables respectively.
%\begin{multline}
%    \label{EQ:MGM_Cond_Cont}
%    p\left(x_{s} \mid x_{/s},y,\Theta\right) = \frac{\sqrt{\beta_{ss}}}{2\pi} * \\ exp\left(\frac{-\beta_{ss}}{2}\left(\frac{\alpha_{s} + \sum_{j}\rho_{sj}\left(y_{j}\right)-\sum_{t \neq s} \beta_{st}x_{t}}{\beta_{ss}} - x_{s}\right)^{2}\right)
%\end{multline}
%
%\begin{multline}
%    \label{EQ:MGM_Cond_Disc}
%       p\left(y_{r} \mid y_{/r},x,\Theta\right) = \\ \frac{exp\left(\sum_{s}\rho_{sr}\left(y_{r}\right)x_{s} + \phi_{rr}\left(y_{r},y_{r}\right) + \sum_{j \neq r} \phi_{rj}\left(y_{r},y_{r}\right)\right)}{\sum_{i=1}^{L_{r}} exp\left(\sum_{s}\rho_{sr}\left(i\right) x_{s} + \phi_{rr}\left(i,i\right) + \sum_{j \neq r} \phi_{rj}\left(i,y_{j}\right)\right)}
%\end{multline}

Learning this model over high dimensional datasets directly is computationally infeasible due to the computation of the partition function, so to avoid this, a proximal gradient method is used to learn a penalized negative log pseudolikelihood form of the model (Equation \ref{EQ:NLL}, product of conditional distributions). To prevent overfitting, nonzero parameters are penalized using the method described in \cite{Sedgewick2016LearningSelection} (Equation \ref{EQ:STEPS}). Here, $\lambda_{CC}$ is a penalty parameter only for edges between continuous variables (CC = Continuous-Continuous), $\lambda_{CD}$ and $\lambda_{DD}$ are for mixed edges and edges only using discrete variables, respectively. $\vert \vert . \vert \vert_{F}$ refers to the Frobenius norm of a matrix. To optimize this objective function the proximal gradient optimization method is used as specified in \cite{Sedgewick2016LearningSelection}.
%\begin{align}
%\begin{split}
\begin{equation}
\label{EQ:NLL}
\widetilde{l}(\Theta\vert x,y) =  -\sum\limits_{s=1}^{p} \log{p(x_{s}\vert x_{\slash s}, y; \Theta)} - \sum\limits_{r=1}^{q}\log{p(y_{r}\vert x,y_{\slash r}; \Theta)}
\end{equation}
%\end{split}
%\end{align}

\begin{equation}
\label{EQ:STEPS}
\underset{\Theta}{\text{minimize}} \, l_{\lambda}(\Theta) = \widetilde{l}(\Theta) + \lambda_{CC} \sum\limits_{s=1}^{p}\sum\limits_{t=1}^{s-1}\vert \beta_{st}\vert + \lambda_{CD} \sum\limits_{s=1}^{p}\sum\limits_{j=1}^{q} \vert \vert \rho_{sj}\vert \vert_{2}  + \lambda_{DD}\sum\limits_{j=1}^{q}\sum\limits_{r=1}^{j-1}\vert \vert \phi_{rj}\vert \vert_{F}
\end{equation}

\section{Methods}
In this section, we describe the computational methods used to realize our pipeline. First, we discuss our variable selection method, and then we discuss how we incorporate prior knowledge to select parameters for this method automatically. Next, we describe the synthetic and real data used to evaluate our approach, and the metrics we apply in our evaluation. Lastly, we describe the prior knowledge sources used.

\subsection{Variable Selection: Preferential Diversity}
The first step in our procedure is to choose variables to model. To this end, we propose to address the following feature selection problem (Definition \ref{DEF:PD}). We first proposed this problem to identify query results relevant to the user but also diverse to give a broad snapshot of the underlying data \cite{PrefDiv}. The problem was referred to as the Top-K relevant and diverse set problem, and is as follows.

\begin{definition}

\label{DEF:PD}
\textbf{Top-K Relevant and Diverse Set}. Given $0 \leq r \leq 1$ a radius of similarity, a set of variables $V$, an output size $k$, a similarity function $Sim\left(V_{i},V_{j}\right)$, and a relevance function $Rel\left(V_{i}\right)$.
\end{definition}

\begin{align}
	\begin{split}
    maximize & \sum_{X_{i}\in S} Rel\left(X_{i}\right)\\ subject\; to &\: S \subset V \\ & \mid S \mid = k \\ & \forall \, i,j \: V_{i} \in S \: and \: V_{j} \in S \rightarrow Sim\left(V_{i},V_{j}\right)<r 
\end{split}
\end{align}

Intuitively, we aim to find a set of variables that are relevant to the user with the constraint that no pair of chosen variables are similar to one another. This is an appropriate choice prior to graphical modeling because graphical models can lose accuracy if redundant variables are included in the model \cite{lemeire2012conservative}. The method we propose to solve this problem is similar in principle to two filter methods: Correlation-based feature selection \cite{hall2000correlation} and maximum relevance minimum redundancy (mRMR) feature selection \cite{ding2005minimum}. Both of these are greedy approaches which select the feature that optimizes an objective function that balances relevance and diversity. The main difference in our approach is that we require zero redundancy, and that we quantify redundancy using prior knowledge. Lastly, to ensure stability of the downstream model, we report the selected features as clusters with redundant variables included in each cluster (instead of discarding them). This allows the user to understand the redundancy in the data.

Another popular approach that follows this principle is the Weighted Gene Correlation Network Analysis (WGCNA) \cite{zhang2005general}. Briefly, this method aims to learn a weighted undirected correlation network by converting correlation to edge weight. With this network, they infer the dissimilarity between nodes in the network, and use network characteristics (e.g. hub nodes) to select important genes. This method differs in that it uses network characteristics instead of summary statistics to infer importance, and the network that they infer is a correlation network whereas graphical models aim to identify conditional dependence relationships. 

Here, we solve this problem using the Preferential Diversity (Pref-Div) algorithm. Pref-Div is an iterative procedure that first selects the top-K most relevant variables and adds them to the result set $R$. Then, it determines whether any pair of variables in $R$ are redundant (as defined by the radius of similarity, $r$ and the similarity function $Sim\left(V_{i},V_{j}\right)$), and removes the lower relevance variable from the result set. Then, the most relevant $K-\mid R\mid$ that have not been explored are added to the result set. This procedure repeats until a set of $K$ relevant and diverse features are selected. For the full procedure, we refer the reader to \cite{PrefDiv}.  In this work, we make one substantial modification from the original Pref-Div algorithm. Here, we compute all variables considered redundant to the selected set of variables and return each variable as a cluster.

We instantiate the Pref-Div algorithm with the following parameter choices. The output size $k$ is user-determined since the appropriate choice for this is based on computational resources available for graphical modeling. Similarity scores between pairs of features are given by pearson correlation, and relevance of each feature is given by pearson correlation to a pre-defined target variable. We note that having a target variable of interest is not necessary, and unsupervised statistics such as variance or domain knowledge can be used to determine relevance scores. In the next section, we discuss how we select the radius of similarity, $\lambda$, using  prior knowledge.

\subsection{Prior Information Pref-Div: piPref-Div}
To choose $\lambda^{*}$, we utilize a method we originally developed to select hyperparameters to learn graphical model structure \cite{manatakis2018pimgm}. Here, we briefly summarize the approach for variable selection: called piPref-Div (Prior Information Pref-Div). The main idea is to compute a correlation graph across many different correlation thresholds, $\lambda^{*}$. A correlation graph contains an edge between $V_{1}$ and $V_{2}$ if the correlation between $V_{1}$ and $V_{2}$ is greater than the correlation threshold.  This method proceeds in four major steps.

\begin{figure*}[!tb]
\begin{subfigure}[]{0.45\textwidth}
\centering
    \includegraphics[width=\textwidth,trim = {6cm 2.5cm 6cm 1.3cm},clip]{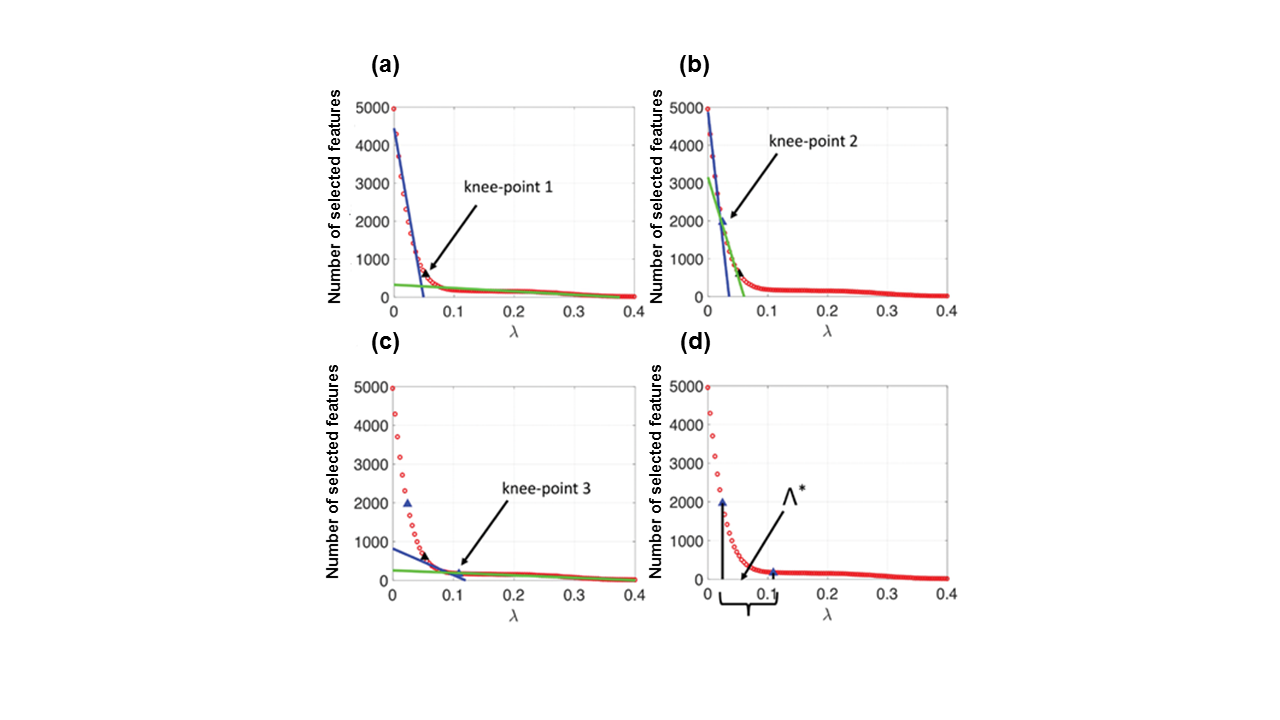}
    \caption{Illustration of procedure to limit tested parameter range.}
    \label{FIG:Knee_Point}
\end{subfigure}
\hfill
\begin{subfigure}[]{0.45\textwidth}
\centering
    \includegraphics[width=\textwidth,trim = {1cm 2.5cm 1cm 2.5cm},clip]{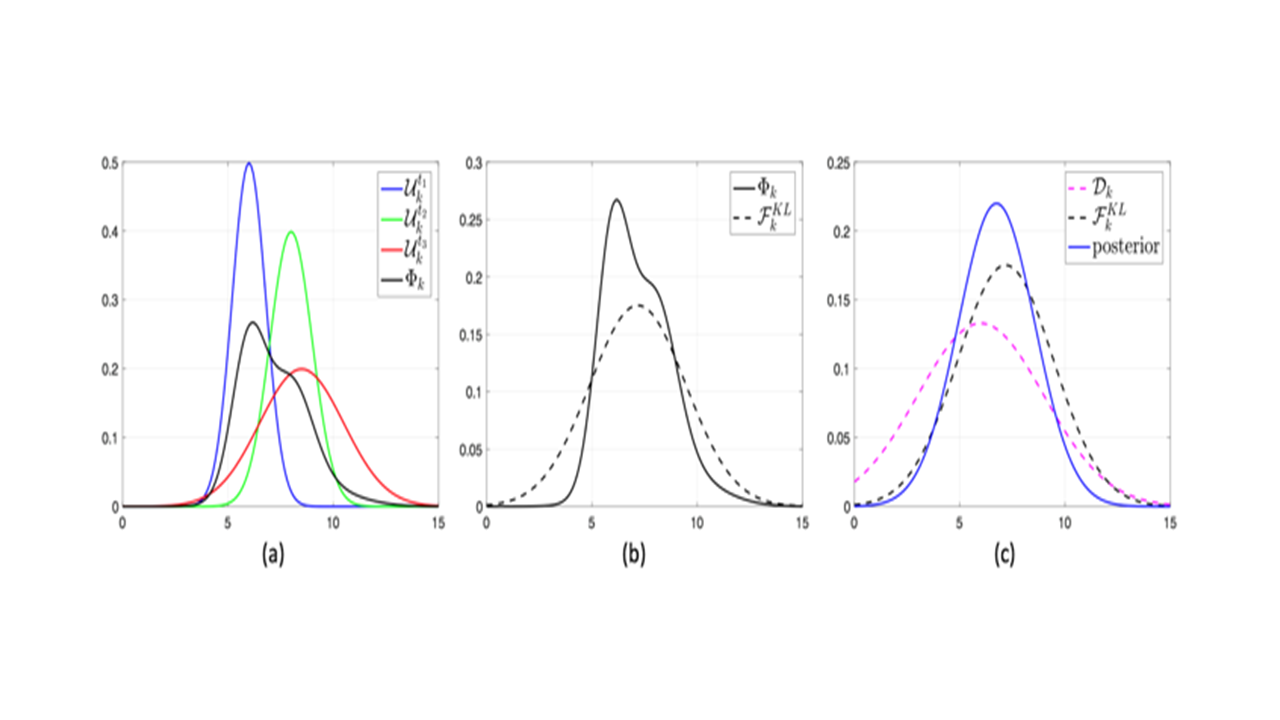}
    \caption{Subsampling procedure to determine empirical probabilities for every edge in the correlation graph. $B\left(\lambda,S\right)$ returns a correlation graph computed upon dataset $S$ with threshold $\lambda$.}
    \label{FIG:Posterior}
\end{subfigure}
\end{figure*}

First, an appropriate parameter range is determined. This is done by identifying a range where the fewest edges are selected (correlation graph is sparse) yet changing $\lambda$ slightly results in a large change in the number of edges in the graph. Figure \ref{FIG:Knee_Point} shows a plot of the number of edges in the correlation graph vs. $\lambda$. Initially, a knee point is identified that best splits the curve into two straight lines (Panel a). Then, this procedure is repeated on each partition of the curve to compute two additional knee-points (Panels b and c). The final parameter range is the space between these two knee-points (Panel d).

%\begin{figure*}[!tb]
%\begin(subfigure)
%\end{subfigure}

%\end{figure*}

Then, a subsampling approach is used to compute empirical probabilities of appearance for each edge by computing correlation graphs across the chosen range of thresholds and random subsamples without replacement. The empirical probability of each edge is its frequency of appearance.

 Next, the information contained in the prior knowledge sources are evaluated against these empirical probabilities across all edges (Equation \ref{EQ:Prior_Eval}). Each prior information source ($t_{r}$) gives knowledge in the form of a probability of appearance for some fixed set of edges ($wp^{t_{r}}$). $\tau_{t_{r}}$ quantifies the "unreliability" of source $t_{r}$. $\phi_{k}^{t_{r}}$ is the expected number of times edge $k$ should appear during the subsampling procedure according to source $t_{r}$, and $\mu_{k}$ is the actual number of appearances for edge $k$.

\begin{equation}
    \label{EQ:Prior_Eval}
    \tau_{t_{r}} = \frac{\sum_{k=1}^{\mid wp^{t_{r}} \mid} \mid \phi_{k}^{t_{r}} - \mu_{k} \mid }{\mid wp^{t_{r}} \mid}
\end{equation}

Finally, posterior distributions are computed for each edge (Figure \ref{FIG:Posterior}). For each edge $k$, a normal distribution is used to approximate the probability of appearance for each prior source (red, green, and blue curves in Panel a). Using a normalized reciprocal of the scores computed in the previous step, these normal distributions are combined into a weighted mixture (black curve, Panel a). This mixture distribution is approximated by the normal distribution which has minimal KL-divergence to the mixture (Panel b). Finally, this normal distribution is combined with a normal distribution from the empirical probabilities to get a posterior distribution (Panel c, blue curve). 

Since some edges may not have prior information from any of the sources, $\lambda^{*}$ is split into two parameters: one for edges where prior information is available $\lambda^{*}_{wp}$ and one for edges where it is not, $\lambda^{*}_{np}$. $\lambda^{*}_{wp}$ is chosen based upon stability of the correlation graph across subsamples along with concordance to the posterior distribution for each feature. $\lambda^{*}_{np}$ is chosen the same way, except that the posterior distribution is the one computed from the data alone (pink curve, Panel c).

\subsection{Simulated Datasets}
\begin{figure}[!tb]
	\centering
     \caption{Cluster Simulation to generate simulated datasets. Purple nodes are master regulators of a cluster, blue nodes are causal parents of the target variable, and the beige node is the target variable. }
    \label{Fig:PD_Simulation}
    \centering
    \includegraphics[width=0.8\linewidth]{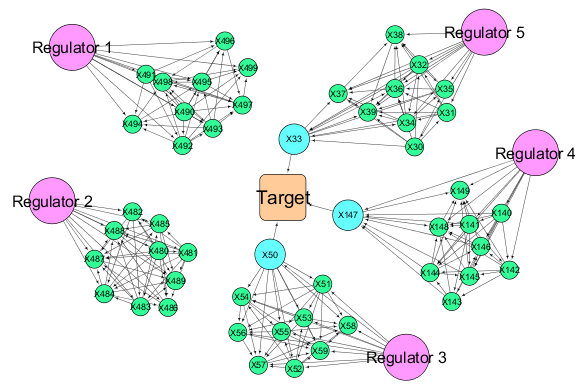}
\end{figure}

Simulated datasets were used to ensure algorithmic correctness and to understand the impact of prior information sources. Data was generated from a linear Gaussian graphical model. Edge coefficients were drawn uniformly at random from the set $\left[-1.5,-0.5\right] \cup \left[0.5,1.5\right]$. Error terms for each variable were zero mean with variance randomly drawn from the set $\left[0.01,2\right]$. Graphical structure was simulated using a "clustered simulation" (Figure \ref{Fig:PD_Simulation}). Here, each variable belonged to one of $C$ clusters. In these clusters, each pair of variables in the cluster were connected by an edge. $c < C$ clusters had one randomly chosen variable connected to the target variable (relevant clusters), while the remaining $C-c$ clusters were disconnected from the rest of the network. Each cluster consisted of an equal number of variables. To represent a master regulator and force correlated structure, each cluster had a single latent (unmeasured) variable that influenced the value of all variables in the cluster. 

Prior knowledge was simulated for two types of prior information sources: reliable priors and unreliable priors. All prior sources give information based on a beta distribution; however, the parameters of this distribution differ based on the type of prior and whether the variables in question belong to the same cluster. An unreliable prior gives information drawn from $Beta\left(4,4\right)$ for both true and false edges (cluster memberships), whereas a reliable prior draws from $Beta\left(10,2\right)$ for true edges, and $Beta\left(2,10\right)$ for false edges. The amount of prior information varies based on the experiment. To determine whether prior information is available for each edge, each edge gets a value $b \backsim U\left(0,1\right)$, and each prior information source has a value $c \in \left[0,1\right]$. The prior gives information about the relationship if $b < c$. In this way, the simulated data reflects the fact that some relationships are more well-studied than others. 

We evaluate piPref-Div on its ability to incorporate unreliable prior information in order to select relevant clusters more accurately. The metric we use for evaluation of selected clusters on simulated data is called cluster accuracy. The goal of this metric is to compare the relevant clusters output by piPref-Div to the truly relevant clusters in the data generating graph, where a relevant cluster is a cluster with at least one variable that is a parent of the target variable. First, an optimal matching is computed between the predicted and actual clusters using the Hungarian Algorithm. The cost of assigning a predicted cluster to an actual cluster is given by 1 - the Jaccard similarity between the clusters. If multiple predicted clusters are best assigned to the same actual cluster, these clusters are combined. Finally, the average Jaccard similarity between the combined predicted clusters and their matched actual clusters are computed as the score.

\subsection{Gene Expression Datasets}

To evaluate the performance of piPref-Div, we apply it to six publicly available breast cancer Affymetrix HGU133A microarray datasets. These datasets have been used in several previous analyses and represent a baseline to evaluate prediction methods \cite{cun2012prognostic,staiger2013current,haury2011influence}. Each dataset consists of microarray expression data for between 159 and 286 patients, and the target variable of interest in this data was whether or not the patient had relapse free survival (RFS)l for 5 years for four of the datasets. For two sets, this information was unavailable and metastasis free survival (MFS) was used (Schmidt et al. and Ivshina et al.). Our evaluation consists of a five-fold cross validation within each dataset to determine how well the selected features give predictive models that are generalizable (accurate) and stable. To measure accuracy, we use area under the ROC curve (AUC) comparing the probability predictions from each method and true binary outcome of RFS and MFS for five years. To measure stability, we use the average tanimoto set similarity (intersection divided by union) for the set of features selected in each fold.

To evaluate the potential of our full pipeline to discover knowledge from data , a graphical model was learned from the TCGA-BRCA RNA-Seq expression dataset. This data included gene expression measurements from 784  breast tumor samples and 13,994 genes. Breast cancer sub-type information for each tumor sample was obtained from \cite{jiang2016comprehensive}. Breast cancer diagnosis and prognosis are commonly divided into four main subtypes: Luminal A, Luminal B, HER2+, and Triple-Negative breast cancer. The main driving distinction for these subtypes is the presence or absence of hormone receptors on the tumor cell surface, which can lead to varying prognoses. In these experiments, we aim to identify clusters distinguishing the four sub-types from expression data. To determine stability of each of these clusters, a 10-fold cross validation was done, and the stability of each cluster was the number of times a similar cluster (similarity > 0.85) was selected in each fold. 

\subsection{External Prior Knowledge Sources}

Prior knowledge consisted of five distinct sources of information. Physical gene distance explored the base pair distance between two genes on the chromosome. If two genes were on separate chromosomes, then this value was set to zero. Otherwise given gene $G_{i}$ from base pairs $B_{i}^{1}$ to $B_{i}^{2}$ and gene $G_{j}$ from base pairs $B_{j}^{1}$ to $B_{j}^{2}$, and full chromosome length $C$, the physical distance prior is given by Equation \ref{EQ:Phys_Dist}. This represents the proportion of chromosome distance covered by the space between these two genes. 
\begin{equation}
    \label{EQ:Phys_Dist}
    Phys\left(G_{i},G_{j}\right) = 1 - \frac{max\left(B_{i}^{2},B_{j}^{2}\right) - min\left(B_{i}^{1},B_{j}^{1}\right)}{C}
\end{equation}
Gene family information was curated from the Human Genome Organization (HUGO). Gene families are groups of genes related by sequence and/or function. A single gene can belong to multiple gene families. Thus, we represent each gene as a vector of families with one-hot encoding. To compute the similarity between these vectors, we use the Jaccard similarity metric which is the number of families in common divided by the total number of unique families either gene belongs to. 
A similar approach is used for gene-disease mapping from the DisGeNet \cite{pinero2016disgenet}. This database gives scores quantifying the level of knowledge that a change in a gene is related to a disease. We use the guilt by association principle to compute whether two genes are related. We represent a gene by a vector of scores to the diseases in the database, and we compute the cosine similarity between two gene vectors. Since all scores are positive, this metric is positive, and is used directly as a probability. 

Finally, we use gene-gene similarity data from two sources: Harmonizome \cite{rouillard2016harmonizome} and STRING \cite{szklarczyk2014string}. Harmonizome similarity data was curated from the Molecular Signatures Database \cite{subramanian2005gene} and consisted of correlation between gene expression across several microarray experiments. STRINGdb curates gene-gene relationship scores based on several factors such as: co-expression, literature co-occurrence, experimental evidence, other databases, etc. STRINGdb scores were scaled from their $\left(0,1000\right)$ range to $\left(0,1\right)$.

\section{Results}
Next, we demonstrate the performance of piPref-Div on simulated and real datasets. First, we evaluate its ability to determine reliable prior information sources and incorporate those sources to select better clusters. Next, we evaluate the method in terms of its ability to accurately predict outcome for breast cancer patients, and lastly, we use the full pipeline to learn a graphical model of breast cancer subtype discrimination.

\subsection{Evaluation and Impact of Prior Knowledge}

\begin{figure}[!tb]
    \centering
    \includegraphics[trim={3.5cm 0.5cm 3.5cm 0cm},width=\columnwidth,clip]{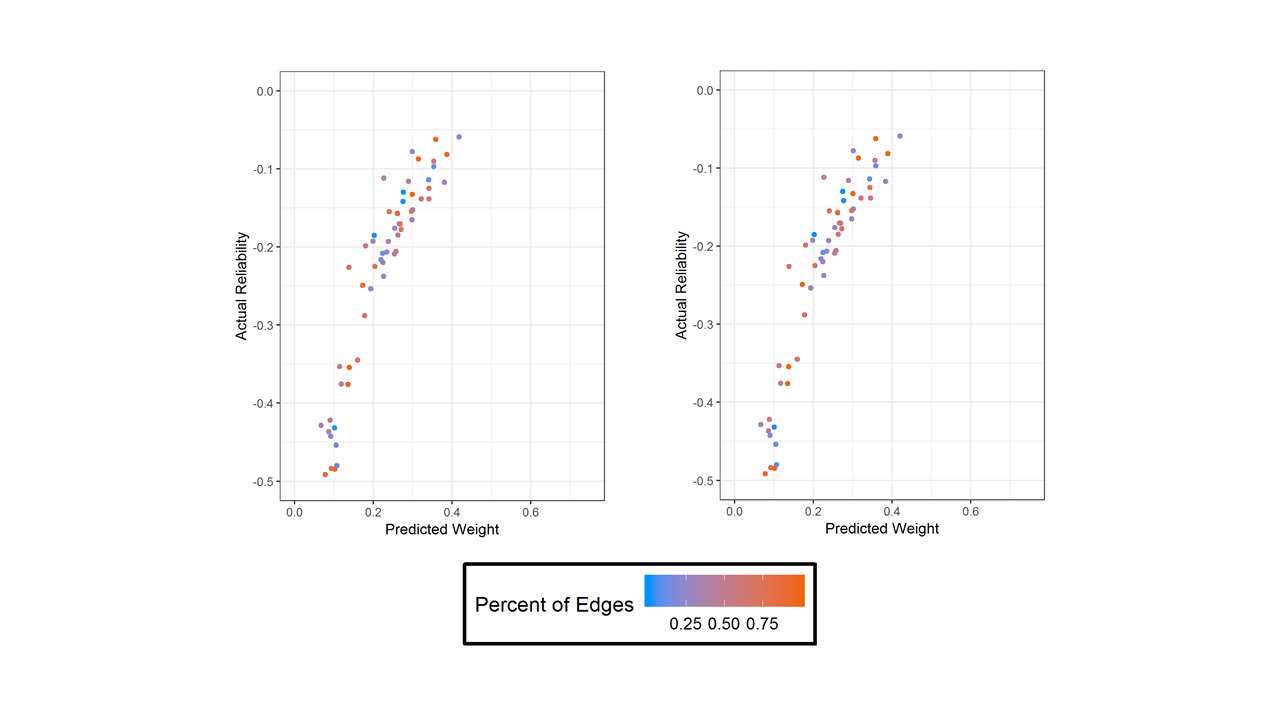}
     \caption{Predicted Weight vs. True Reliability for each prior knowledge source in simulated experiments for piPref-Div for (left) 50 samples and (right) 200 samples.}
    \label{Fig:Prior_Evaluation_PD}
\end{figure}

First, we tested the ability of piPref-Div to accurately evaluate prior knowledge sources on simulated datasets of 500 variables with 50 clusters (25 relevant), 200 and 50 samples, and 5 prior knowledge sources (3 reliable) with a random amount of prior information. The results are presented in Figure \ref{Fig:Prior_Evaluation_PD}. Here the "Actual Reliability" on the y-axis refers to the sum of the probabilities given to true edges minus the sum of the probabilities given to false edges for each prior. The predicted weight for each prior knowledge source given by piPref-Div shows a clear association to the reliability score. A benefit of this approach is that this weight does not appear to be dependent on the amount of prior information. Even with little prior information (blue circles), piPref-Div assigns an accurate weight to the knowledge sources. 

The next experiment investigated the impact of the amount and quality of prior knowledge on the ability for piPref-Div to identify relevant clusters of variables in the data. In these experiments, we test the method using the same experimental parameters as the prior knowledge evaluation section, and using a larger dataset with 3000 variables, 300 clusters (75 relevant). For each experimental setting, 15 graphs were generated and the results are presented cumulatively over these graphs.

\begin{figure}[!tb]
\begin{subfigure}[b]{0.48\columnwidth}
    \centering
    \includegraphics[trim={0cm 0cm 0cm 0.75cm},width=\linewidth]{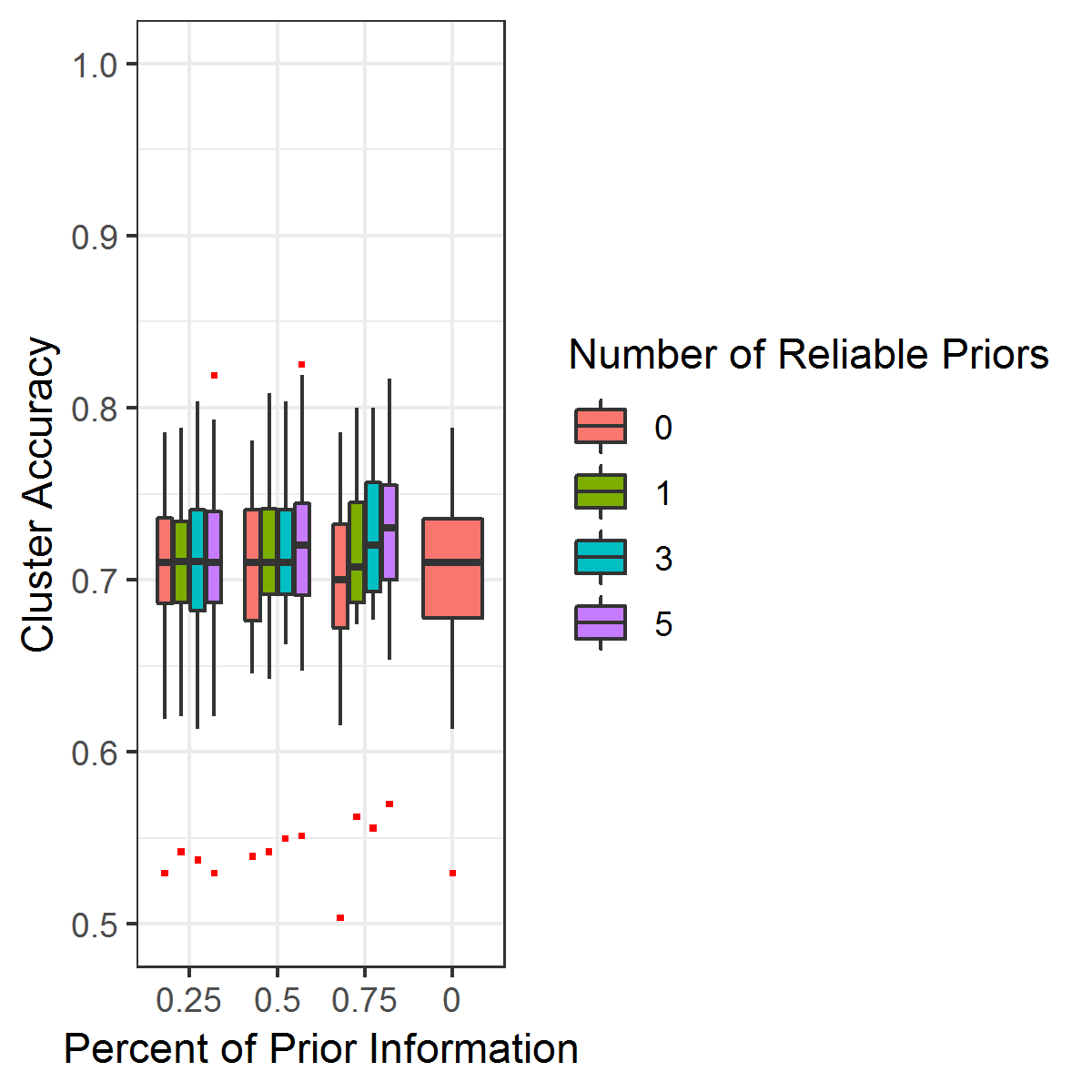}
    \end{subfigure}
      \hfill
    \begin{subfigure}[b]{0.48\columnwidth}
        \centering
        \includegraphics[trim={0cm 0cm 0cm 0.75cm},width=\linewidth]{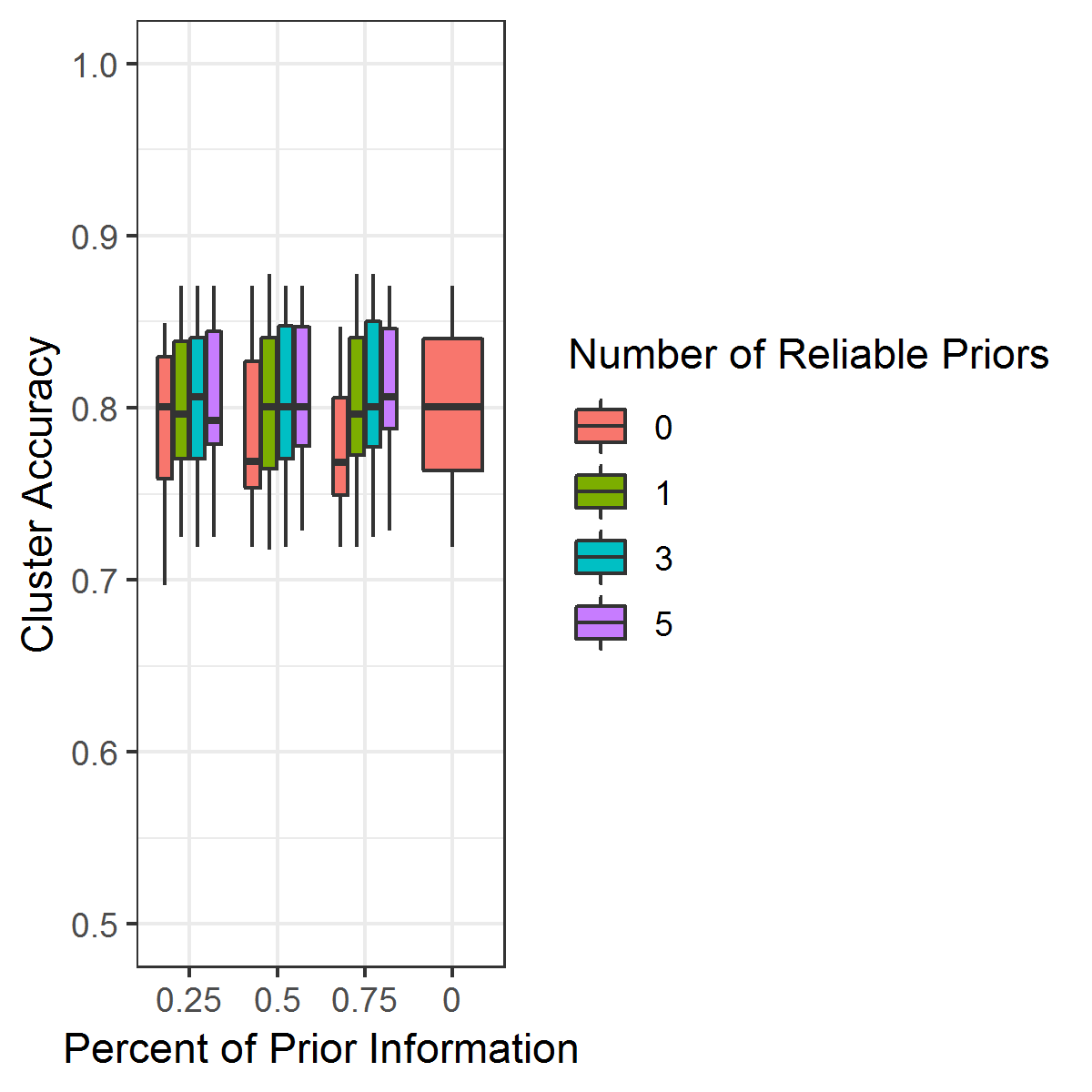}
    \end{subfigure}
     \caption{Accuracy of predicted clusters for varying amount and reliability of prior knowledge. Sample size was set to 50 (left) and 200 (right).}
    \label{Fig:PD_Small_Cluster}
\end{figure}

The results for the small datasets are given by Figure \ref{Fig:PD_Small_Cluster}. Sample size is clearly the most significant factor in determining accuracy of the selected clusters. Prior information gives a modest improvement in accuracy, but this benefit only occurs with at least 50\% of prior information and at least 3 reliable sources out of 5. However, when all sources are unreliable, there is no decrease in accuracy unless there is a large amount of information present. Lastly, we note that the benefit of prior information is drastically reduced in cases with sufficient sample size (200 sample case). This is intuitive, as with more data, correlation becomes a very stable measure, and prior information can be ignored.

\begin{figure}[!tb]
\begin{subfigure}[b]{0.48\columnwidth}
    \centering
    \includegraphics[trim={0cm 0cm 0cm 0.75cm},width=\textwidth]{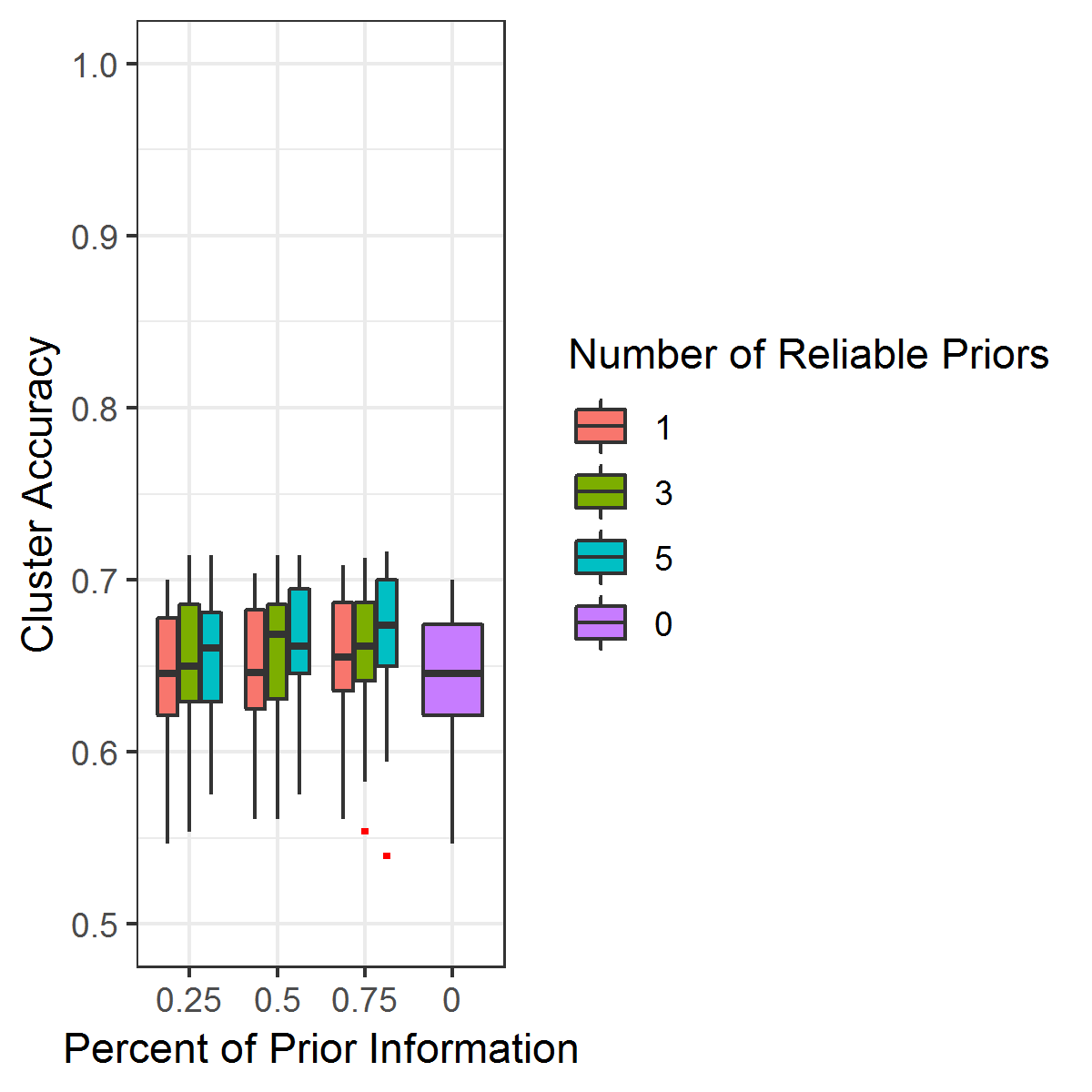}
    \end{subfigure}
      \hfill
    \begin{subfigure}[b]{0.48\columnwidth}
        \centering
        \includegraphics[trim={0cm 0cm 0cm 0.75cm},width=\textwidth]{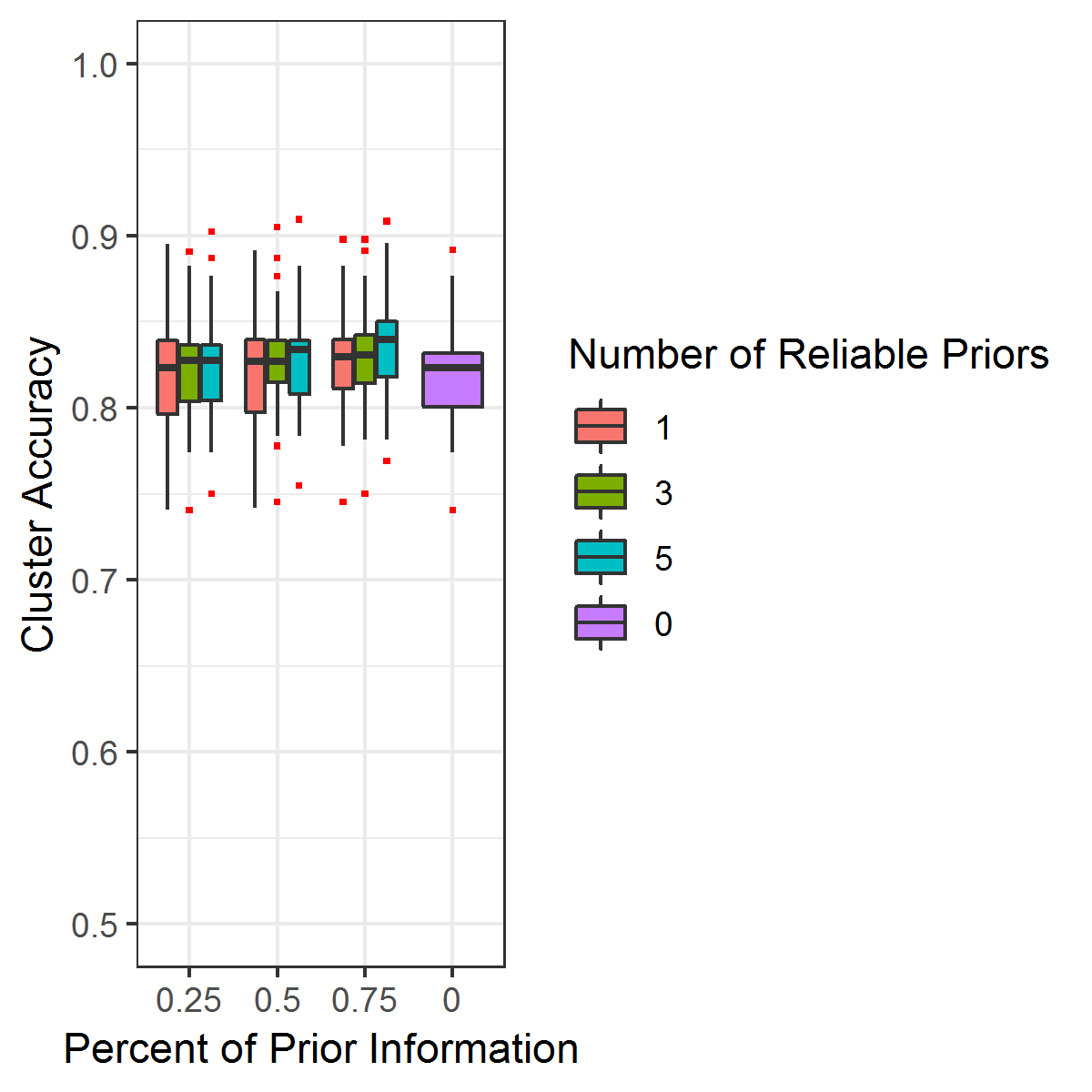}
    \end{subfigure}
     \caption{Accuracy of predicted clusters for varying amount and reliability of prior knowledge on large datasets. Sample size was set to 50 (left) and 200 (right).}
    \label{Fig:PD_Large_Clust}
\end{figure}

Lastly, we examined the ability of piPref-Div to detect clusters from a larger graph (Figure \ref{Fig:PD_Large_Clust}). Here, the pattern is largely similar, except the impact of prior knowledge is more significant. In particular, even 25\% prior information gives a substantial increase in accuracy over having no prior information at all. Again, this impact is larger when the sample size is small, though it is present in both cases. Further, the impact of more samples is more pronounced in the larger dataset. An increase from 50 to 200 samples results in an increase in accuracy from 0.65 to over 0.8 for all amounts of prior.

\subsection{Breast Cancer Outcome Prediction}

To determine the performance of piPref-Div on real datasets, we applied the algorithm to the aforementioned breast cancer microarray datasets. Three variations of piPref-Div were tested. piPref-Div alone (PD), piPref-Div with and without prior information (No Prior =  NP) with clusters aggregated into summarized features using principal component analysis (PD-PCA,PDNP-PCA). For the Pref-Div approaches, an inner 3-fold cross-validation loop was used to determine the number of selected features (1,3,5, and 10 features were tested). Genes with less than 0.5 standard deviation across samples in the training set were removed from the dataset prior to feature selection. For comparison purposes, two methods that performed well in a previous study were included in the analysis: Hybrid-Huberized SVM (HH-SVM) and Recursive-Reweighted Feature Elimination (RRFE) \cite{cun2012prognostic}.

\begin{figure*}[!tb]
    \centering
    \includegraphics[trim={0cm 0.5cm 0cm 0cm},clip]{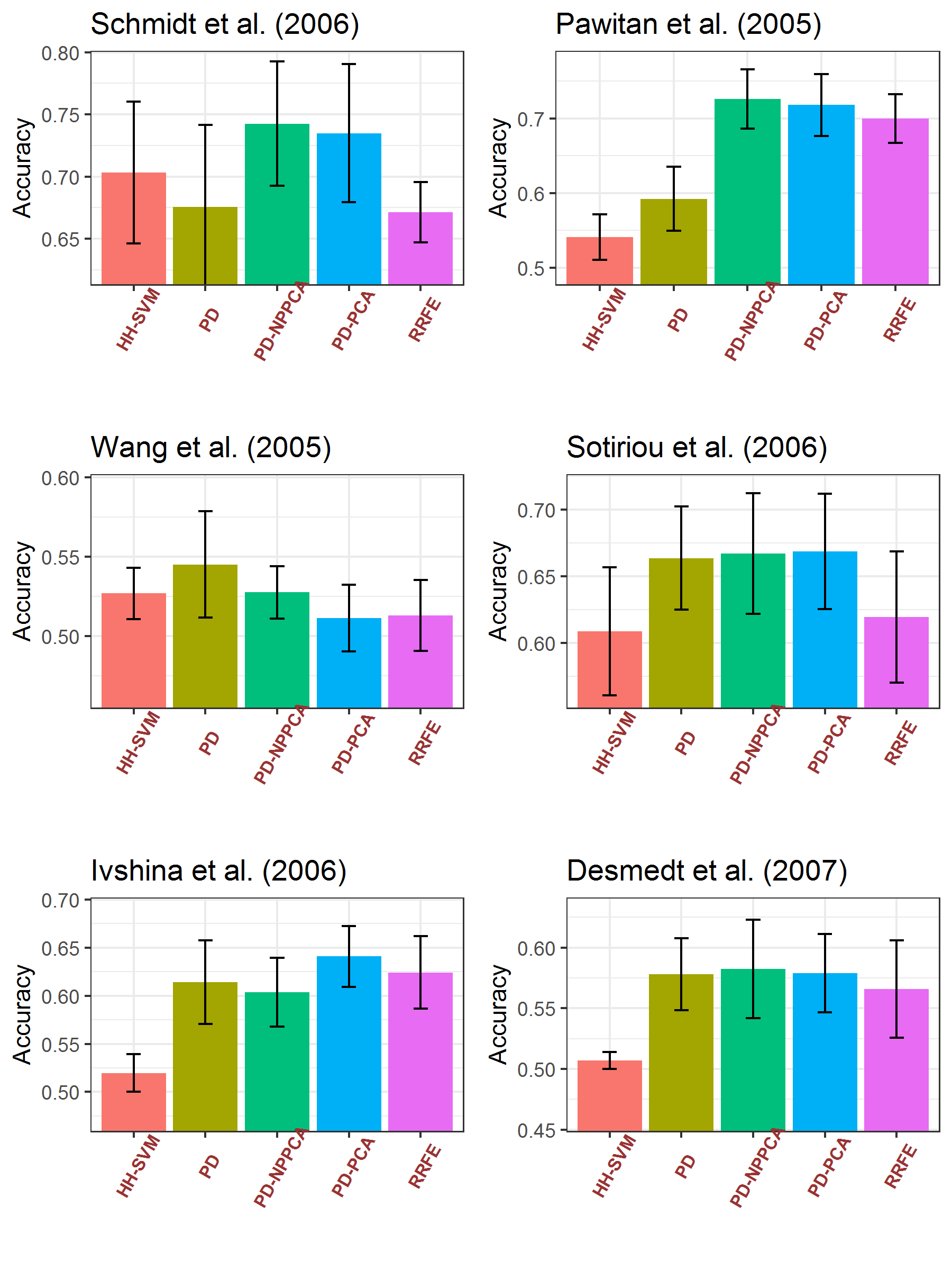}
    \caption{AUC of predicting RFS using several feature selection methods on six independent breast cancer microarray datasets.}
    \label{Fig:AUC_Variable_Selection_Results}
\end{figure*}

Figure \ref{Fig:AUC_Variable_Selection_Results} presents the accuracy results. Across the datasets, the consistent best performing methods are PD-NP-PCA and PD-PCA (sea-green and light blue, respectively). This suggests that cluster selection and representing individual features as clusters offers a benefit to selecting single genes alone. However, this is dataset dependent, as Pref-Div alone (PD, yellow box) matches these methods on 2 of the datasets (Sotiriou and Desmedt) and performs better on 1 dataset (Wang). Overall, these results show no significant difference between using prior information (PD, PD-PCA) and not using prior information (PD-NPPCA). Using prior information with PCA clustering shows a slight improvement on the Ivshina dataset, but none of the others. We find that our approach performs about the same in terms of accuracy when compared to SVM-RRFE, but our method tends to select significantly fewer features.

Figure \ref{Fig:Stability_Variable_Selection_Results} presents the stability of the learned models. The results confirm previous work that identifying a stable model for breast cancer outcome prediction is a difficult problem \cite{cun2012prognostic}. In general, only the RRFE algorithm shows somewhat consistent stability; however, we note that a major contributing factor is that this algorithm uses on average 119 selected features, whereas HH-SVM averages around 6 and the PD approaches average around 1 feature (or cluster). Among, the Pref-Div based approaches, PD-PCA with and without prior information show the most consistent stability. On nearly all datasets they are near to, or on par with RRFE despite choosing significantly fewer features.

\subsection{Stratification of Breast Cancer Subtypes}
Finally, we evaluate our full pipeline of variable selection and then graphical modeling based on its ability to mine interesting clusters related to breast cancer subtype. Based on the previous section, we chose to use PD-PCA for variable selection due to its consistently high accuracy and relatively high stability. An MGM model was learned on a dataset consisting of only the selected clusters and the Subtype variable.  To summarize clusters into single names, the Ingenuity Pathway Analysis regulator analysis was used, and the KEGG Pathway database was queried (corrected p-values < 0.05 were chosen as candidates). Following this step, only specific pathways and regulators were included as names of the clusters. 

\begin{figure}[!tb]
\centering
\includegraphics[trim={3cm 0cm 5cm 0cm},clip,width=0.8\columnwidth]{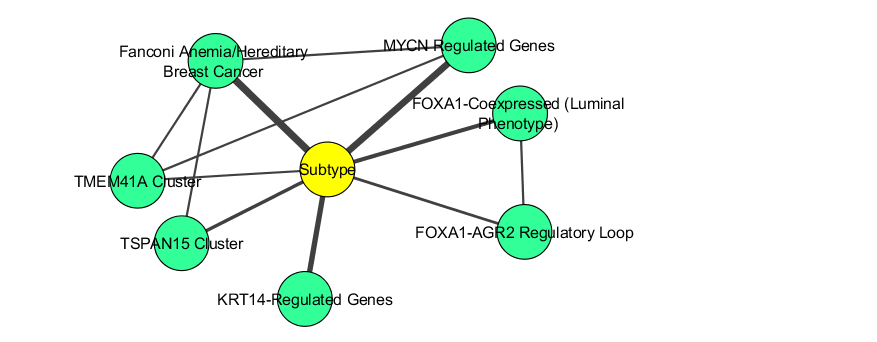}
\caption{Graphical model of breast cancer subtype. Size of each edge represents the number of times a similar cluster was selected to be related to Subtype in each of the cross-validation folds.}
\label{FIG:Subtype_Graph}
\end{figure}

The learned graphical model is presented in Figure \ref{FIG:Subtype_Graph}. We found that two clusters were unable to be mapped coherently to any biological function (single gene representatives were TMEM41A, and TSPAN15); however, these clusters were relatively unstable. The two most stable clusters were: Fanconi Anemia/ Hereditary Breast Cancer pathway, and a set of genes regulated by MYCN.  Fanconi Anemia and the Hereditary Breast Cancer pathways are known to share common genes \cite{alan2010fanconi}, and developing breast cancer through a genetic basis tends to be associated with ER+ breast cancer \cite{mulligan2011common}. MYC family pathways and the transcription factors themselves are known to be differentially expressed across subtypes, and the MYCN factor in particular has shown differences between triple-negative and other subtypes \cite{horiuchi2012myc}. FOXA1 along with GATA3 and ESR1 are necessary for maintaing a luminal phenotype of breast cancer \cite{chaudhary2017novel}, and AGR2 is upregulated by FOXA1 but only in an estrogen receptor dependent manner \cite{wright2014delineation}. This implies that the FOXA1-AGR2 loop will only be upregulated in ER+ breast cancer. Though it is unclear how KRT14 regulated genes distinguish subtypes of breast cancer, it is known that upregulation of KRT14 reduces the ability of breast tumor to metastasize and invade the extracellular matrix \cite{westcott2015epigenetically}. Overall, we find that the pipeline constructs and selects reasonable clusters that are discriminative of breast cancer subtypes. In addition, the pipeline generates novel candidate clusters for scientists to experimentally probe.

\section{Discussion}

In this work, we  have presented a pipeline to learn graphical model structure from large omics datasets. The pipeline builds upon previous work in integrating and evaluating prior information to select hyperparameters, and a variable selection method to identify relevant yet non-redundant sets of features. We have extended this approach to return clusters of variables instead of individual features, and to use these features as input to a graphical model structure learning algorithm to find interesting relationships within the clusters and between clusters and phenotypes. 

We evaluated our work on both synthetic data and real breast cancer data. On the synthetic data, we find that our method is able to accurately evaluate prior information, and utilize this information to improve the selection of relevant clusters. In addition, the method avoids performing more poorly than having no prior information at all, when most of the prior information available is unreliable. We find that the largest improvement over having no prior information occurs when there are few samples and a large number of features. Overall, we find the improvement when using prior information to be modest, but this appears to be necessary to avoid poor performance with unreliable priors.

On classification experiments with microarray data, we find that our method performs at par or better with other state of the art approaches. We find that using PCA to summarize clusters into single variables gives a slight improvement over selecting single variables alone. We find that when compared to the state of the art approaches, our method selects far fewer features to achieve similar or better levels of accuracy. When using the full pipeline with graphical modeling to analyze RNA-Seq data to discriminate breast cancer subtypes, we find that the method identifies reasonable clusters. Two of the seven clusters did not map to any known biological regulator or pathway, and are candidates for further investigation. 

For future work, we aim to improve upon the accuracy results with reliable prior information. It could be that using the prior information solely to select hyperparameters is too conservative an approach, and using the posterior distributions directly can give better results. Since the priors are already being appropriately weighted, the posterior distributions should be accurate. In addition, we aim to utilize our pipeline with experimental approaches to validate selected clusters in terms of activity and biological significnce. Finally, the prior information sources used for the biological experiments were relatively sparse. It is future work to utilize the vast array of gene expression experiments available to construct priors that cover a wider representation of the genome.

%\section{Appendices}
%
%If your work needs an appendix, add it before the ``\verb|\end{document}|'' command at the conclusion of your source document. 
%
%Start the appendix with the ``\verb|appendix|'' command:
%\begin{verbatim}
%  \appendix
%\end{verbatim}
%and note that in the appendix, sections are lettered, not numbered. This document has two appendices, demonstrating the section and subsection identification method.
%
%\section{SIGCHI Extended Abstracts}
%
%The ``\verb|sigchi-a|'' template style (available only in \LaTeX\ and not in Word) produces a landscape-orientation formatted article, with a wide left margin. Three environments are available for use with the ``\verb|sigchi-a|'' template style, and produce formatted output in the margin:
%\begin{itemize}
%\item {\verb|sidebar|}:  Place formatted text in the margin.
%\item {\verb|marginfigure|}: Place a figure in the margin.
%\item {\verb|margintable|}: Place a table in the margin.
%\end{itemize}

%
% The acknowledgments section is defined using the "acks" environment (and NOT an unnumbered section). This ensures
% the proper identification of the section in the article metadata, and the consistent spelling of the heading.
\section{Acknowledgements}
This work was supported by NIH Grants U01HL137159, R01LM012087 (to PVB), and T32CA082084 (to VKR).

%%TODO

%
% The next two lines define the bibliography style to be used, and the bibliography file.
\bibliographystyle{plain}
\bibliography{document}

% 
% If your work has an appendix, this is the place to put it.
\newpage
\appendix

\section{Supporting Results}

\begin{figure*}[!b]
    \centering
    \includegraphics[trim={0cm 1cm 0cm 0cm},width=0.8\textwidth,clip]{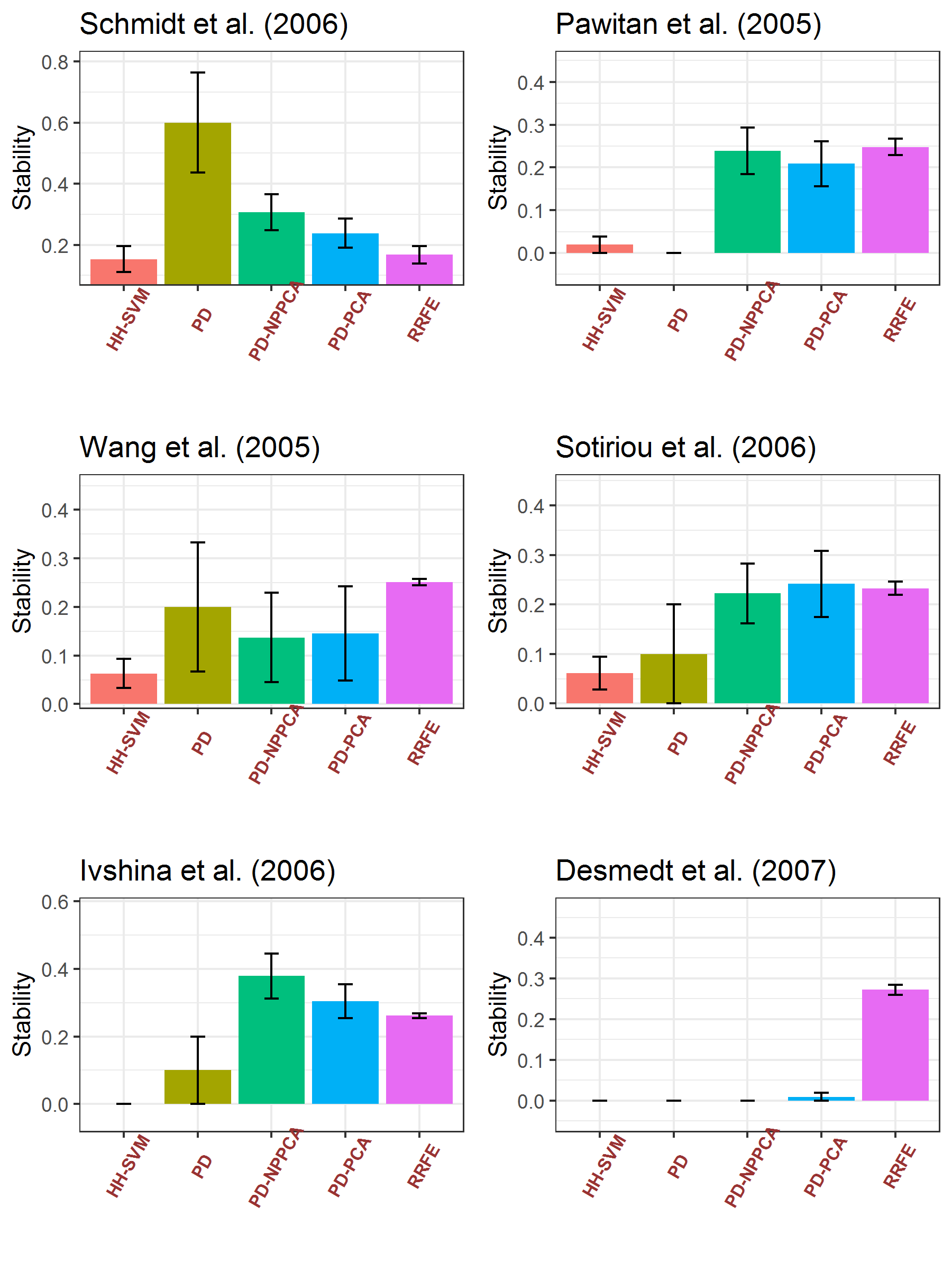}
    \caption{Stability of learned models for predicting RFS using several feature selection methods on six independent breast cancer microarray datasets.}
    \label{Fig:Stability_Variable_Selection_Results}
\end{figure*}

%
%\section{Research Methods}
%
%\subsection{Part One}
%
%Lorem ipsum dolor sit amet, consectetur adipiscing elit. Morbi malesuada, quam in pulvinar varius, metus nunc fermentum urna, id sollicitudin purus odio sit amet enim. Aliquam ullamcorper eu ipsum vel mollis. Curabitur quis dictum nisl. Phasellus vel semper risus, et lacinia dolor. Integer ultricies commodo sem nec semper. 
%
%\subsection{Part Two}
%
%Etiam commodo feugiat nisl pulvinar pellentesque. Etiam auctor sodales ligula, non varius nibh pulvinar semper. Suspendisse nec lectus non ipsum convallis congue hendrerit vitae sapien. Donec at laoreet eros. Vivamus non purus placerat, scelerisque diam eu, cursus ante. Etiam aliquam tortor auctor efficitur mattis. 

\end{document}